\documentclass[aps,prl,twocolumn,showpacs,showkeys,nofootinbib,superscriptaddress]{revtex4-1}

\usepackage[utf8]{inputenc}
\usepackage{graphicx}
\usepackage{amsfonts}
\usepackage{amssymb}
\usepackage{amsmath}
\usepackage[mathscr]{eucal}
\usepackage{setspace}
\usepackage{booktabs}
\usepackage[shellescape,latex]{gmp}
\def\sqr#1#2{{\vcenter{\hrule height.#2pt
   \hbox{\vrule width.#2pt height#1pt \kern#1pt
      \vrule width.#2pt}
   \hrule height.#2pt}}}

\def\bsqr#1#2{{\vrule width #1pt height#2pt}}
\def\bsquare{{\mathchoice\bsqr66\bsqr66\bsqr33\bsqr33}}
\def\badbreak{\penalty1000}

\def\union{\cup}                    % union
\newcommand{\cS}{{\cal S}}                  % cal-S
\newcommand{\cL}{{\cal L}}                  % cal-L
\def\fir{{\scriptscriptstyle{\text{\rm IR}}}}                 % IR subscript 
\def\fuv{{\scriptscriptstyle{\text{\rm UV}}}}              % UV subscript 
\def\fa{{\scriptscriptstyle{\text{\rm A}}}}                  % Anderson subscript
\def\bin{{\scriptscriptstyle{\text{\rm B}}}}                 % B subscript
\def\lm0{{\lambda_0}}                                             % IR scale 1
\def\nrN{N}                                                       %  number of objects
\def\efN{\mathscr{N}}                                        %  effective number of objects
\def\efNm{\efN_\star}                                        %  minimal effective number of objects
\def\v{b}                                                               %  counting weight
\makeatletter
\newcommand*{\smallrel}[2][.8]{%
  \mathrel{\mathpalette{\smallrel@{#1}}{#2}}%
}
\newcommand*{\smallrel@}[3]{%
  \sbox0{$#2\vcenter{}$}%
  \dimen@=\ht0 %
  \raise\dimen@\hbox{%
    \scalebox{#1}{%
      \raise-\dimen@\hbox{$#2#3\m@th$}%
    }%
  }%
}
\makeatother

\usepackage{amsmath} % wonderful math package
\usepackage{dsfont}  % double strike font
\usepackage{bm}      % bold math

\def\beq{\begin{equation}}
\def\eeq{\end{equation}}
\def\beqs#1\eeqs{\beq\begin{split} #1 \end{split}\eeq}

\long\def\comment#1{}

\usepackage{microtype}
\usepackage[dvipsnames]{xcolor}
\usepackage[colorlinks=true,backref=false, linktocpage=true,
citecolor=PineGreen,urlcolor=PineGreen,linkcolor=PineGreen,pdfpagemode=UseOutlines]{hyperref}

\hypersetup{%
  bookmarksnumbered=true,
  pdftitle = {},
  pdfsubject = {},
  pdfauthor = {},
  pdfkeywords = {}
}

\begin{document}

\title{Localized Modes in the IR Phase of QCD}

\author{Andrei\ Alexandru}
\email{aalexan@gwu.edu}
\affiliation{The George Washington University, Washington, DC 20052, USA}

\author{Ivan Horv\'ath}
\email{ihorv2@g.uky.edu}
\affiliation{University of Kentucky, Lexington, KY 40506, USA}
\affiliation{Nuclear Physics Institute CAS, 25068 \v{R}e\v{z} (Prague), Czech Republic}

\author{Neel Bhattacharyya}
\email{neel468167@gmail.com}
\affiliation{Poolesville High School, 17501 W Willard Rd, Poolesville, MD 20837, USA}

\date{Dec 5, 2023}

\begin{abstract}
Infrared (IR) dimension function $d_\fir(\lambda)$ characterizes the space effectively 
utilized by QCD quarks at Dirac scale $\lambda$, and indirectly the space occupied
by glue fields. It was proposed that its non-analytic behavior in thermal 
{\em infrared phase} reflects the separation of QCD system into an IR component and an 
independent bulk. 
Here we study the ``plateau modes" in IR component, whose dimensional properties were 
puzzling. Indeeed, in the recent {\em metal-to-critical} scenario of transition to IR 
phase, this low-dimensional plateau connects the Anderson-like mobility edge
$\lambda_\fir \!=\!0$ in Dirac spectrum with mobility edges $\pm \lambda_\fa$. 
For this structure to be truly Anderson-like, plateau modes have to be exponentially 
localized, implying that both the effective distances $L_\text{eff} \propto L^\gamma$ 
and the effective volumes $V_\text{eff} \propto L^{d_\fir}$ in these modes grow slower 
than any positive power of IR cutoff $L$. Although $\gamma\!=\!0$ was confirmed in 
the plateau, it was found that $d_\fir \!\approx\! 1$. Here we apply the recently 
proposed {\em multidimension} technique to the problem. We conclude that a plateau 
mode of pure-glue QCD at UV cutoff $a \!=\! 0.085\,$fm occupies a subvolume of IR 
dimension zero with probability at least 0.9999, substantiating this aspect 
of metal-to-critical scenario to a respective degree.

\medskip

\end{abstract}

\maketitle

%\section{Introduction.}

\noindent
{\bf 1.~Introduction. $\,$}
Recent developments in thermal~QCD \cite{Alexandru:2019gdm, Alexandru:2021pap, Alexandru:2021xoi}, 
enabled by studies of lattice-regularized systems, led to a remarkable alignment of two 
seemingly unrelated aspects: the recently proposed infrared (IR) phase~\cite{Alexandru:2019gdm}, 
and the older suggestion of Anderson-like localization in Dirac 
spectra~\cite{GarciaGarcia:2005vj, GarciaGarcia:2006gr, Kovacs:2010wx, Giordano:2013taa}.

\begin{figure}[t]
\begin{center}
    \vskip -0.0in
    \centerline{
    \hskip 0.0in
    \includegraphics[width=8.0truecm,angle=0]{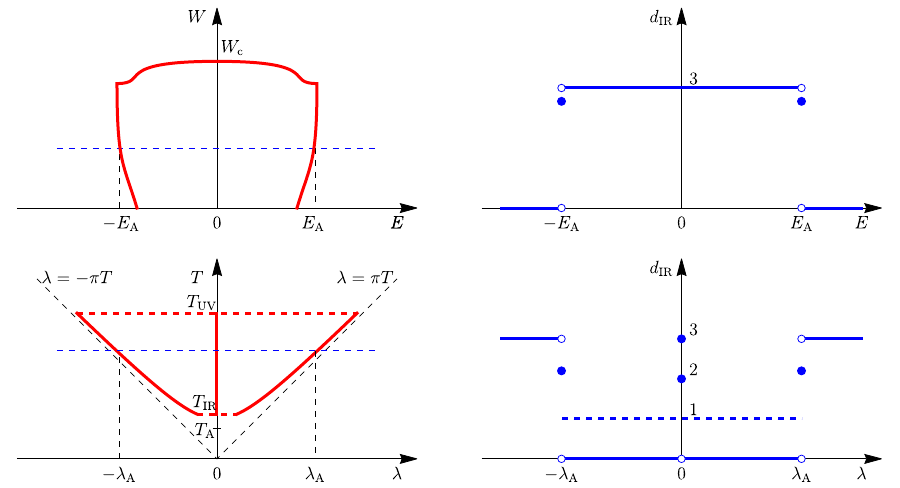}
    }
    \vskip -0.05in
    \caption{Phase diagrams of localization in Anderson models (top) and 
    QCD in IR phase (bottom). In Anderson case, $E_A$ is the critical energy 
    at given disorder strength $W$. In QCD case~\cite{Alexandru:2021xoi}, $T_\fuv$ 
    denotes the possible endpoint of IR phase (onset of perturbative regime), 
    and $T_A$ the crossover temperature, both defined in Ref.~\cite{Alexandru:2019gdm}.
    See text for other explanations.}
    \label{fig:Dirac_diagram}
    \vskip -0.50in
\end{center}
\end{figure}

This fusion arose via two new elements. The first one is the notion of 
{\em effective counting dimension}~\cite{Horvath:2022ewv,Horvath:2018aap,Horvath:2018xgx}
characterizing probability measures constructed via discrete regularizations. 
This was adopted as a tool to describe IR and UV properties of quantum states or Dirac modes 
in terms of spatial dimensions $d_\fir$ and $d_\fuv$ respectively~\cite{Alexandru:2021pap}.
The second element is the proposal of Ref.~\cite{Alexandru:2021xoi}, 
{\em metal-to-critical scenario}, that QCD in IR phase features in fact two types 
of Anderson-like mobility edges: in addition to previously known $\pm \lambda_\fa$ 
($\lambda_\fa \!>\! 0$)~\cite{Kovacs:2010wx, Giordano:2013taa}, there is also a strictly 
IR mobility edge at $\lambda_\fir \!=\!0$~\cite{Alexandru:2021xoi}. 
To visualize the situation, top left panel in Fig.~\ref{fig:Dirac_diagram} shows the standard 
phase diagram of Anderson models in $E \!-\! W$ (energy -- disorder strength) plane. Here 
the region enclosed by red solid line of critical points contains extended states. The QCD 
analogue is in the bottom left panel, showing the critical lines in $\lambda \!-\! T$ 
(Dirac eigenvalue -- temperature) plane~\cite{Alexandru:2021xoi}. Note the extra 
line $\lambda_\fir(T) \equiv 0 \,,\, T_\fir \!\le\! T \!\le\! T_\fuv$, where 
$T_\fir$ marks the transition to IR phase. The region enclosed by critical lines 
contains localized modes in this case. Hence, the relationship between QCD and Anderson 
situations is of dual rather than direct nature.

The rationale for $d_\fir$ is that it properly characterizes the physical space 
effectively occupied by electrons/quarks in various regimes. 
Indeed, it is based on effective counting~\cite{Horvath:2018xgx} which is 
additive and thus defines meaningful spatial volumes (measures). This is not 
the case for generalized dimensions, such as frequently used $D_2$.
Dimension $d_\fir$ also conveniently 
identifies critical lines as collections of points where $d_\fir(E,W)$ or 
$d_\fir(\lambda,T)$ are non-analytic. The $(E,W)$ setup defines critical features 
of Anderson localization phenomenon~\cite{Anderson:1958a,Anderson:1980a}, and 
known aspects of $d_\fir$ in this case are as 
follows~\cite{Horvath:2021zjk,Horvath:2022lbj,Horvath:2022klk}. In extended 
regime, $d_\fir \!=\!3$ is expected, which was confirmed numerically to very high 
accuracy~\cite{Horvath:2021zjk}. It was also determined that $d_\fir \!\approx\! 8/3$ 
at Anderson criticality irrespective of the universality class~\cite{Horvath:2021zjk}. 
The value of $d_\fir$ and the~degree of superuniversality are claimed to about 
2-3 parts per mill.\footnote{The expression for $d_\fir$ in terms of 
multifractal spectrum has recently been proposed and slightly higher value 
$d_\fir \!=\!2.733(3)$ (orthogonal class) based on it was 
suggested~\cite{Bur2023B} but questioned in~\cite{Hor2023B, Horvath:2022com}.}
Localized wave functions are expected to be bounded by a decaying exponential, 
which yields $d_\fir \!=\!0$. While this last aspect is not fully 
guaranteed (exponential localization is not rigorously proved in 3 dimensions),
the top right panel of Fig.~\ref{fig:Dirac_diagram} shows the present-day
picture of generic $d_\fir(E)$ for~$0 \!<\! W \!<\! W_c$. 

%and numerical studies of $d_\fir$ in this regime have not been performed (see below). 

For Dirac spectrum in IR phase of QCD to be truly Anderson-like, the values of 
$d_\fir$ in corresponding regimes need to match the above. Conversely, the nature 
of differences, if any, should be clarified. In the bottom right panel of 
Fig.~\ref{fig:Dirac_diagram} we show the generic $d_\fir(\lambda)$ for 
$T_\fir \!<\! T \!<\! T_\fuv$ obtained in the pure-glue QCD analysis
of Ref.~\cite{Alexandru:2021pap}. In the presumed localized regime 
$(-\lambda_A,0) \union (0,\lambda_A)$, $d_\fir \!\approx\! 1$ was found 
(dotted line) instead of the expected $d_\fir \!\approx\! 0$.
Here we resolve this discrepancy. In Sec.~2 we perform the first direct calculation 
of $d_\fir$ in the Anderson localized regime and compare it to that in QCD. 
Similar behaviors are found with very slow decrease of $d_\fir$ estimates toward 
$L \!\to\! \infty$ limit. We attribute this to the logarithmic growth of effective 
volume for pure exponential which yields $d_\fir \!=\!0$ but slow 
$L \!\to\! \infty$ convergence. The observed QCD-Anderson similarities then 
suggest that $d_\fir \!=\!0$ in both~cases. In Sec.~3 we perform the first 
{\em multidimensional analysis}~\cite{Horvath:2022klk} of both Anderson-localized 
states and QCD-localized Dirac modes, which makes a clear case that the two 
are indeed dimensionally equivalent.

\noindent
{\bf 2. Direct Evaluation.} IR dimension $d_\fir$ is a leading power of linear 
size $L$ ($L \!\to\! \infty$) in average effective volume 
$\langle \,\efNm[\psi]\, \rangle_{L,\lambda} \,\propto\, L^{d_\fir(\lambda)}$.
Here $\psi$ denotes Dirac eigenmodes $D \psi(x) \!=\! i\lambda \psi(x)$ or Anderson 
states at given energy.  A useful concept is the finite-volume IR 
dimension~\cite{Horvath:2021zjk}
\begin{equation}
  d_\fir(L,s) \equiv \frac{1}{\ln(s)} 
  \ln \frac{\langle \,\efNm\rangle_L}{\langle\, \efNm\rangle_{L/s}} 
  \;\; , \;\, \lim_{L \to \infty} d_\fir(L,s) \!= d_\fir
  \label{eq:40}     
\end{equation}  
with any $0 < s \neq 1$. If $P \!=\! (p_1,p_2,\ldots,p_\nrN)$, 
$p_i \!\equiv\! \psi^\dagger\psi(x_i)$ are probabilities entailed by $\psi$,
then~\cite{Horvath:2018aap} 
\begin{equation}
     \efNm[\psi] \,\equiv\, \efNm[P] \,=\, 
     \sum_{i=1}^\nrN \min\, \{ \nrN p_i, 1 \}.    \;\,
     \label{eq:080}              
\end{equation}

\begin{figure}[b]
\begin{center}
    \vskip -0.23in
    \centerline{
    \hskip -0.02in
    \includegraphics[width=4.5truecm,angle=0]{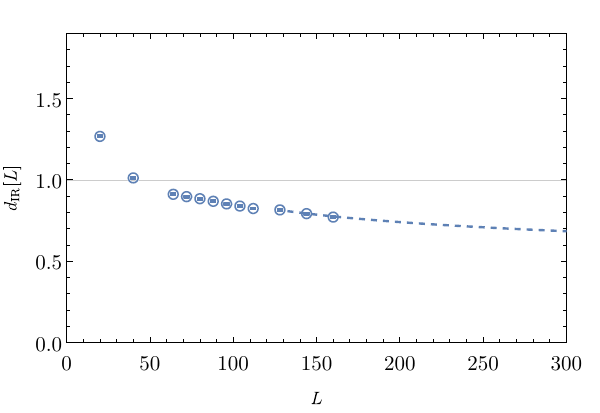}
    \hskip -0.12in
    \includegraphics[width=4.5truecm,angle=0]{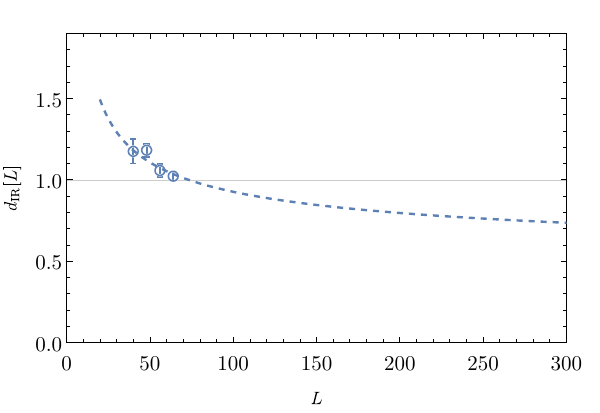}    
    }
    \vskip -0.10in
    \caption{Function $d_\fir(L) \!\equiv\! d_\fir(L,2)$ in the Anderson model
    (O-class) at $W \!=\! 32$ (left panel) and ``plateau modes" in IR phase of 
    pure-glue QCD. The dashed line, used to guide the eye, corresponds to a fit 
    of the form $\langle \,\efNm\rangle_L\sim[\log(L)]^k$. The effective size 
    $\ell$~\cite{Alexandru:2021xoi} of modes is $\ell \!\approx\! 2.9$ in 
    the Anderson and $\ell \!\approx\! 5.2$ in the QCD case.}
    \label{fig:direct}
    \vskip -0.40in
\end{center}
\end{figure}

As a first step in making the case for $d_\fir \!=\! 0$ of QCD plateau modes 
we compare their $d_\fir(L,s)$ to that in a generic Anderson model. 
The Hamiltonian of the latter is (O class, periodic boundary conditions) 
\begin{equation}
     {\cal H} \,=\, \sum_r \epsilon_r \, c^\dag_r  \, c_r 
     \,+\, \sum_{r,j}  c^\dag_r \, c_{r-e_j} + h.c.
     \label{eq:065}
\end{equation}
where $r\!=\!(x_1,x_2,x_3)$ are lattice sites, $e_j$ ($j\!=\!1,2,3$) unit 
lattice vectors, $\epsilon_r \!\in\! [-W/2,+W/2]$ uniformly distributed 
random potentials, and $c_r$ the electron operators. We focus on 1-particle
states in the vicinity of zero $E$ (energy) at $W \!=\!32$, which is deeply
in the localized regime~\cite{Markos:2006}. Note that $\langle \ldots \rangle$ in Eq.~\eqref{eq:40} 
refers to disorder average in this case. For QCD it is the path-integral 
average. 

In Fig.~\ref{fig:direct} we show results for $d_\fir(L) \!\equiv\! d_\fir(L,s\!=\!2)$. 
The left panel features Anderson data at $L$ up to $160$. JADAMILU 
package~\cite{jadamilu_2007} was used to perform the numerical diagonalization. 
The right panel shows the analogous data at available volumes of pure-glue QCD 
in the IR-phase setup of Ref.~\cite{Alexandru:2021pap},
i.e. $T\!=\!1.12\, T_\fir$, Wilson action at $\beta \!=\! 6.054$ 
($a\!=\!0.085\,$fm, $r_0\!=\!0.5\,$fm), $N_t \!=\! 1/(Ta) \!=\! 7$.  
Eigenmodes of the overlap operator ($\rho \!=\! 26/19$) were computed and analyzed on 
systems with sizes up to $L \!=\! 72$. Numerical implementation is described 
in Refs.~\cite{Alexandru:2011sc, Alexandru:2014mqy, Alexandru:2011ee}. 
We used modes in the eigenvalue range $\lambda \in (150,450)\,$MeV which is safely 
inside the plateau region~\cite{Alexandru:2021pap}. 

%$L/a \!=\!24,28,32,40,48,56,64$ and $1/(Ta)\!=\!7$. 
%The implementation of overlap is described 
%in Refs.~\cite{Alexandru:2011sc, Alexandru:2014mqy, Alexandru:2011ee}.
%Additional technical details can be found 
%in Refs.~\cite{Alexandru:2019gdm, Alexandru:2021pap}.
%Overlap discretization is crucial to distinguish zero modes from the nearby modes. 
%On our largest volumes the low-lying modes have unusually small 
%eigenvalues and we numerically diagonalize the overlap operator $D$
%rather than more conventional $D^\dagger D$. Using this approach,
%we find that low-lying modes are separated from zero modes by 
%several orders of magnitude. 

Fig.~\ref{fig:direct} reveals similar dimensional behaviors in the two cases, 
with $d_\fir(L) \!\approx\! 1$ in the region of accessible QCD sizes, and 
subsequent slow decrease toward the infrared in the Anderson case. 
Hence, the two nominally very different dynamics can indeed have a common 
Anderson-localization origin. Since there is little doubt that 
$d_\fir \!=\!0$ in Anderson case, one expects the same in QCD.
Also, for decaying exponential of width $\sigma$, it is easy to show 
that the leading $L \!\to\! \infty$ term in $\efNm(L)$ is proportional 
to $\log^3(L/\sigma)$. It is this type of behavior that likely causes 
the slow decrease of $d_\fir(L)$.

\noindent
{\bf 3. Multidimensional Analysis.} While the above makes $d_\fir \!=\!0$ 
for QCD plateau modes in IR phase plausible, convincing demonstration of 
dimensional equivalence to localized Anderson states is essential. We will 
show that the recently developed 
{\em multidimensional analysis}~\cite{Horvath:2022klk} (MDA) provides 
what is needed. 

MDA aims to resolve the dimensional substructure (if any) in a probability 
measure defined by a discrete (e.g. lattice) regularization. 
It is conceptually different from effective counting dimensions 
(e.g. $d_\fir$): rather than defining an {\em effective subset} of sample 
space and specifying its dimension, MDA considers a family of ordinary 
fixed subsets containing points with similar probabilities. Scaling 
properties of their volumes can reveal the presence of distinct dimensions. 
MDA differs from multifractal formalism~\cite{Halsey_Kadanoff_multif:1986} 
in that it focuses on physically relevant populations, namely 
those whose volumes contain non-zero total probability 
in $L \!\to\! \infty$ limit (See also Ref.~\cite{Horvath:2022com}.). 

Given Anderson states $\psi$, MDA first orders probabilities in vectors $P[\psi]$ 
via $p_1 \!\ge p_2 \!\ge \ldots \ge p_{N(L)}$. Closeness within $P$ then 
generically entails closeness of probabilities, and ``populations" are defined 
by suitable sequential segments in $P$. To that end, vector 
$(q_0,q_1,\ldots,q_{\nrN})$ of cumulative probabilities is formed, namely
$q_0 \!=\! 0$, $q_j \!=\! q_{j-1} + p(j)$, and function $\nu(q)$ of cumulative 
counts, namely $\nu(0) \!=\!0$, $\nu(1) \!=\! \nrN$, 
$\nu(q) = j(q) + (q-q_j)/(q_{j+1}-q_j)$ for $0 \!<\! q < 1$, is constructed.
Here $j(q)$, $q \!\in\! (0,1)$ is the largest $j$ such that $q_j \!<\! q$.
Clearly, $\nu(q)$ is the number of spatial points with largest probabilities,
summing up to $q$. Their collection $\cS(q)$ is a subset of lattice space.
MDA then defines
\begin{equation}
    d(q) := \dim \cS(q) \quad \text{i.e.} 
    \quad \nu(q,L) \propto L^{d(q)}
    \;\; \text{for} \;\;
    L \!\to\! \infty \,
    \label{eq:125}    
\end{equation}

Given the order in $P$, $\nu(q)$ is increasing and convex, and $d(q)$ is
non-decreasing~\cite{Horvath:2022klk}. Hence, $d(q)$ also arises ``differentially" 
as dimension of $\sigma(q,\epsilon) \!:=\! \cS(q) \setminus \cS(q-\epsilon)$
\footnote{Note that $d(q)$ represents the largest dimension present in $\cS(q)$.}
\begin{equation}
    \nu(q,L) - \nu(q-\epsilon,L) \propto L^{d(q,\epsilon)}
    \quad , \quad  d(q,\epsilon)=d(q) 
    \label{eq:165}    
\end{equation}
for all $0 < \epsilon < q$. This allows for
{\em dimensional decomposition} of lattice space $\cL$, and well-defined 
occurrence probabilities for all dimensions~\cite{Horvath:2022klk}. To that end, 
MDA partitions interval $[0,1]$ into $B$ equal parts of width 
$\epsilon_\bin \!=\! 1/B$, thus defining $B$-tuple of $q$-values 
$q_b \!=\! b/B$, $b=1,\ldots,B$. 
Then $\cL(L) = \cup_{b=1}^{B} \sigma(q_b,\epsilon_\bin,L)$ and
\begin{equation}
   \nrN(L) = \sum_{b=1}^{B} v(q_b,\epsilon_\bin,L) \, L^{d(q_b,\epsilon_\bin,L)}
   \label{eq:205}   
\end{equation}
with finite-volume $d$ and $v$ introduced via~\cite{Horvath:2021zjk,Horvath:2022klk}
\begin{equation}
   d(q,\epsilon,L) = \frac{1}{\log s} \,
   \log \frac{\nu(q,L) - \nu(q-\epsilon,L)}{\nu(q,L/s) - \nu(q-\epsilon,L/s)} 
   \;\;\;    
   \label{eq:245}    
\end{equation}
($0 \!<\! s \!\neq\! 1$) and 
$v(q,\epsilon,L) L^{d(q,\epsilon,L)} = \nu(q,L) - \nu(q-\epsilon,L)$. 
Relation \eqref{eq:205} is exact at each $B$, and defines formal expressions 
such as $\nrN(L) \!=\! \int_0^1 dq \,v(q,L)L^{d(q,L)}$. In a setup with suitable 
$B$ (thus $\epsilon_\bin$), Eq.~\eqref{eq:245} will be used here to estimate 
$d(q)$ since $\lim_{L\to \infty} d(q,\epsilon,L) = d(q)$ for all 
$0 \!<\!\epsilon \!<\!q$. Probability $p$ of dimension $d$ 
is $p(d)\!=\! \int_0^1 dq\, \delta(d-d(q))$~\cite{Horvath:2022klk}.

We described MDA using a sequence of states $\psi \!=\!\psi(L)$ 
labeled by increasing $L$, and the associated cumulative counts $\nu(q,L)$. 
However, all MDA calculations presented here are based on 
$\nu \rightarrow \langle \nu \rangle$, where $\langle \ldots \rangle$
denotes the disorder average in Anderson case and the path-integral average 
in QCD case.

%The occurrence probability $p$ of dimension $d$ (``dimension content" $p(d)$) 
%is $p(d)\!=\! \int_0^1 dq\, \delta(d-d(q))$ \cite{Horvath:2022klk}.
%\begin{equation}
%    \lim_{L\to \infty} d(q,\epsilon,L) = d(q) \;\;,\;\; \forall \, 0< \epsilon < q 
%\end{equation}

%If abundance of population labeled by $q$ scales as ${\cal N}(q,L) \propto L^{d(q)}$, 
%then its sample probabilities have to scale as $\bar{p}(q,L) \propto L^{-d(q)}$ 
%for total probability ${\cal N} \bar{p}$ to remain non-zero. This guides 
%the choice of $q$~\cite{Horvath:2022klk}: if associated $d(q)$ is monotonic, 
%then ordering sample points by magnitude and growing $L$ increasingly better 
%separates the populations~\cite{Horvath:2022klk}.

\begin{figure}[t]
\begin{center}
    \vskip -0.05in
    \centerline{
    \hskip  0.05in
    \includegraphics[width=4.63truecm,angle=0]{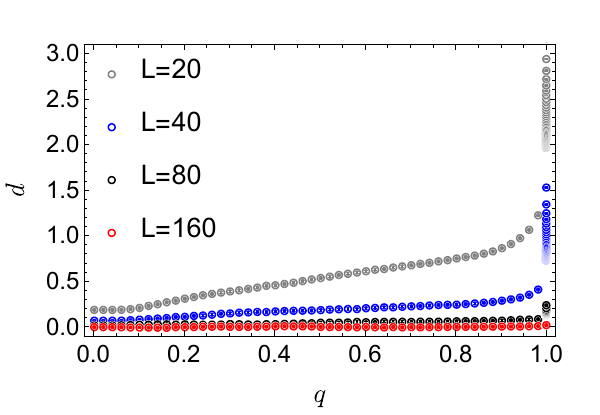}
    \hskip -0.16in
    \includegraphics[width=4.47truecm,angle=0]{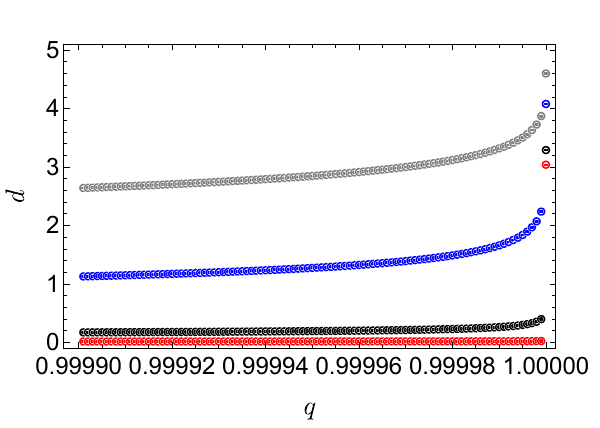}    
    }
    \vskip -0.10in
    \caption{Finite-$L$ $d(q)$ in 3-d Anderson model (O-class) at $W \!=\! 32$
    ($s \!=\! 2$). 
    Left: calculation at $B\!=\!10^3$ (last shown bin is $b\!=\!B\!-\!1$).
    Right: calculation at $B\!=\!10^6$ (last bin is $b\!=\!B$).}
    \label{fig:d_vs_q_And}
    \vskip -0.45in
\end{center}
\end{figure}

\noindent
{\bf 4. Plateau Modes.}
We now perform MDA of Anderson states (localized phase) and of QCD plateau modes 
(IR phase) using setups described in Sec.~2. Starting with the former,
Fig.~\ref{fig:d_vs_q_And} (left) shows $d(q)$ calculated at $B=10^3$ ($s=2$) for 
increasing sizes $L$. Notice that $q$-dependences at each $L$ already exhibit
the monotonicity that is only guaranteed in $L \!\to\! \infty$ limit. From definition 
of $d(q)$ it follows that $d(1) \!=\! 3$. Hence, calculation at given $B$ provides 
non-trivial information about $d(q)$ on the interval $[1/B, 1-1/B]$. Results for 
bin $b \!=\!1$ and $b\!=\!B\!-\!1$ (the first and the last data point shown for 
each $L$) correspond to these endpoint values.
The key observation is that, while non-zero $d(q)$ do appear at finite $L$, they
get quickly reduced as $L$ increases. In the present case they effectively 
vanish at $L\!=\!160$.

The above provides numerical evidence that $d(q) \!=\! 0$ for $q \in (0,0.999]$. 
However, it doesn't exclude the possibility that higher dimensions appear with 
probability smaller than $\epsilon_\bin \!=\!10^{-3}$. Indeed, this information is hidden 
within the last bin, i.e. the $q$-interval $(0.999,1]$. To uncover its behavior, we divided it
into another $10^3$ sub-intervals, thus working at resolution $\epsilon_\bin \!=\!10^{-6}$. 
Results for the right-most edge of $q$ is shown in Fig.~\ref{fig:d_vs_q_And} (right), this 
time also including the last bin. Remarkably, non-zero dimensions again quickly scale out 
to zero for $q \!\le\! 1-10^{-6}$. This is visible directly in $L\!=\!160$ data before
the $L \!\to\! \infty$ extrapolation. Notice that even the $q\!=\!1$ result (last bin) is 
already settled very near the correct value $d(1)\!=\!3$.

\begin{figure}[b]
\begin{center}
    \vskip -0.15in
    \centerline{
    \hskip  0.05in
    \includegraphics[width=3.5truecm,angle=0]{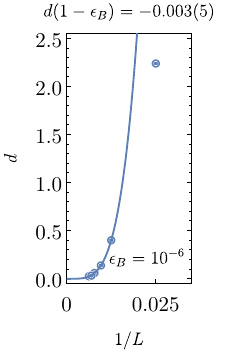}
    \hskip  0.05in
    \includegraphics[width=3.5truecm,angle=0]{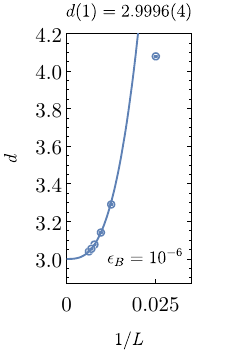}    
    }
    \vskip -0.10in
    \caption{Extrapolation ($L \!\to\! \infty $) of 
    $d(q,\epsilon_\bin,L)$ for $q \!=\! 1 \!-\! \epsilon_\bin$ (left) 
    and $q \!=\! 1$ (right) in 3-d Anderson model (O-class) at $W \!=\! 32$. 
    See text for details.}
    \label{fig:extrapolations}
    \vskip -0.40in
\end{center}
\end{figure}

To address the numerical rigor in the above, we show the $1/L \!\to\! 0$ extrapolations 
for $q \!=\! 1 \!-\! 10^{-6}$ (next to last bin) and $q\!=\!1$ (last bin) in 
Fig.~\ref{fig:extrapolations}. The $L$-dependence was fitted to $d(q)$ plus general
power using 5 largest values of $L$, and had excellent $\chi^2/\text{dof}$.
Note that $d(1)$ for the smallest (not fitted) lattice is significantly larger than
dimension of the underlying space. This occurs because, by construction, the last 
bin has to accommodate all low-d populations whose total probability vanishes in 
thermodynamic limit. Such transport proceeds via a flow of volume toward the last bin 
at finite $L$, and results in an unphysically large finite-L dimension.
Our results (indicated in Fig.~\ref{fig:extrapolations}) substantiate the conclusion 
that,  for $E\!=\!0$ Anderson states at $W \!=\! 32$~(O-class)
\begin{equation}
   p(d) = (1-\Delta) \, \delta(d) + \Delta \, \bar{p}(d)
   \quad ,\quad 
   \Delta < 10^{-6}
   \label{eq:285}    
\end{equation}
In other words, that the probabilty of encountering a non-zero spatial dimension 
(potentially featured in an unknown distribution $\bar{p}(d)$) is smaller than $10^{-6}$.

\begin{figure}[t]
\begin{center}
    \vskip 0.03in
    \centerline{
    \hskip  0.05in
    \includegraphics[width=4.25truecm,angle=0]{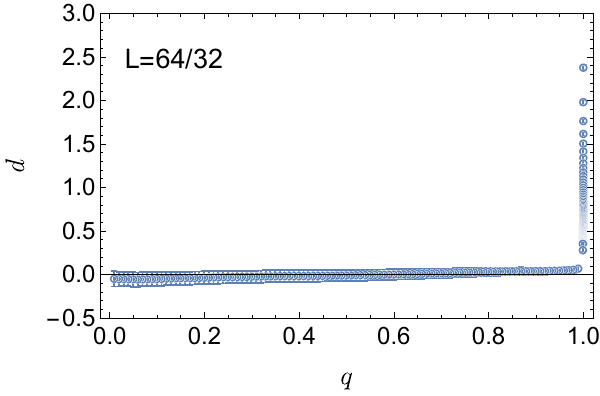}
    \hskip  -0.05in
    \includegraphics[width=4.25truecm,angle=0]{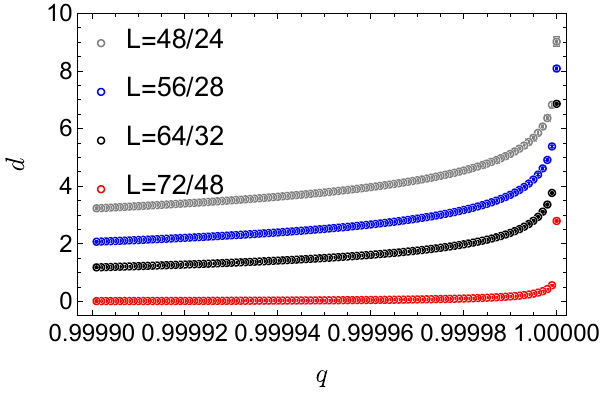}    
    }
    \vskip -0.07in
    \caption{Function $d(q, \epsilon_\bin,L)$ for QCD plateau modes.
    Left: at $B\!=\! 10^3$ and $L\!=\!64$, $s\!=\!2$. 
    Right: at $B\!=\! 10^6$ and indicated pairs of $L$. The last shown 
    bin is always $b \!=\! B\!-\!1$ ($q \!=\! 1 \!-\! \epsilon_\bin$).}
    \label{fig:QCDplateau_d}
    \vskip -0.40in
\end{center}
\end{figure}

Turning now to QCD in IR phase, we again work with plateau modes in the range 
$\lambda \in (150,450)\,$MeV~\cite{Alexandru:2021pap}. 
In Fig.~\ref{fig:QCDplateau_d} (left) we show finite-L result at 
$L \!=\!64$ ($B \!=\! 10^3$). Note that, even without an extrapolation, $d$ 
is consistent with zero in almost the entire $q$-domain. 
To decipher the behavior at the very right edge where $d(q,L)$ rises, we again 
re-analyze the last bin using $B\!=\!10^6$. Results for various pairs of 
sizes are shown in Fig.~\ref{fig:QCDplateau_d} (right). Similarly to the Anderson 
case (Fig.~\ref{fig:d_vs_q_And} (right)), the rise gets quickly reduced as $L$ 
increases, leaving behind yet narrower and weaker rise at $L \!=\! 72$. 

\begin{figure}[t]
\begin{center}
    \vskip 0.02in
    \centerline{
    \hskip  0.05in
    \includegraphics[width=4.33truecm,angle=0]{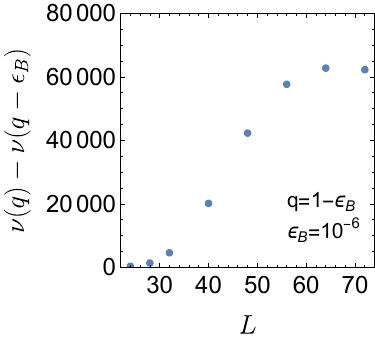}
    \hskip  -0.00in
    \includegraphics[width=4.00truecm,angle=0]{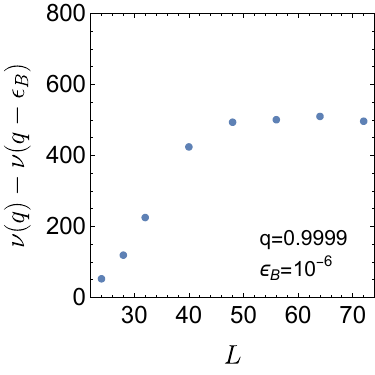}    
    }
    \vskip -0.07in
    \caption{Average counts $\nu(q) \!-\! \nu(q-\epsilon_\bin)$ in QCD 
    plateau modes at $\epsilon_\bin \!=\! 10^{-6}$ for $q=1 \!-\! \epsilon_\bin$ 
    (left) and $q \!=\! 0.9999$ (right).}
    \label{fig:QCDplateau_nu}
    \vskip -0.40in
\end{center}
\end{figure}

Statistical strength and the range of sizes in available QCD data is not 
sufficient to perform numerical analysis at $\epsilon_\bin \!=\! 10^{-6}$ 
in the same way as in Anderson case (see Eq.~\eqref{eq:285}). Nevertheless, 
the point can be made convincingly here as well. To that end, we plot in 
Fig.~\ref{fig:QCDplateau_nu} (left) average counts in the next-to-last bin
($q \!=\! 1 \!-\! \epsilon_\bin$, $\epsilon_\bin \!=\! 10^{-6}$) and observe 
a tentative saturation involving the largest two systems $L \!=\! 64,\,72$. 
True saturation would imply that $d(q) \!=\! 0$ for $q \!\le\! 1-\epsilon_\bin$. 
Whether this is indeed taking place can be checked farther away from 
$q \!=\! 1$ edge, where counts should saturate at smaller $L$.
Fig.~\ref{fig:QCDplateau_nu} (right) shows this for $q\!=\! 0.9999$,
revealing that a wider plateau is indeed formed.  
This leads us to propose that Eq.~\eqref{eq:285} in fact holds also 
in the QCD case. 

\noindent
{\bf 5.~Summary. $\,$} While the notion of IR phase in QCD~\cite{Alexandru:2019gdm} 
was sparked by the IR behavior of Dirac spectral density $\rho(\lambda)$, it is
the unusual effective spatial dimensions of Dirac modes~\cite{Alexandru:2021pap}, 
expressed by spectral function $d_\fir(\lambda)$, that became a key element 
in understanding the phase and detecting it. Indeed, essential attributes 
of IR phase, namely the existence of deep IR 
fields~\cite{Alexandru:2015fxa, Alexandru:2019gdm} and their separation (decoupling) 
from the bulk~\cite{Alexandru:2019gdm}, became natural in {\em metal-to-critical} 
picture of IR transition~\cite{Alexandru:2021xoi}. The underlying mechanism gives 
special role to Anderson-like critical points in Dirac spectra: the pair 
$\pm \lambda_\fa$, $\lambda_\fa \!>\! 0$ facilitates the decoupling, while 
$\lambda_\fir \!=\! 0$ governs the proposed long-range and possibly exactly 
scale-invariant physics of the IR component~\cite{Alexandru:2021xoi, Alexandru:2019gdm}. 
Since Anderson critical points transform the space available to a particle by 
changing its dimension (see~\cite{Horvath:2021zjk,Horvath:2022lbj,Horvath:2022klk}), 
metal-to-critical scenario entails a specific discontinuous $d_\fir(\lambda)$: 
a blueprint of IR phase. 

However, evidence that 
produced~\cite{Alexandru:2019gdm, Alexandru:2021pap, Alexandru:2021xoi} and later
corroborated~\cite{Meng:2023nxf} the metal-to-critical scenario also generated 
an inconsistency. Indeed, the mechanism requires that plateau modes 
of IR phase (e.g. $\lambda_\fir \!<\! \lambda \!<\! \lambda_\fa$) are exponentially 
localized, and thus of zero IR dimension. But the numerical evidence pointed 
toward $d_\fir \!\approx\! 1$ instead. Here we resolved this issue by direct 
confrontation of Anderson-model data and QCD data, both in terms of $d_\fir$ 
and the new multidimensional technique function $d(q)$.
Our analysis leaves little doubt that Anderson localized states and QCD plateau
modes are dimensionally equivalent, both behaving as spatial probabilistic 
objects of IR dimension zero. This removes the above inconsistency. 

\begin{figure}[b]
\begin{center}
    \vskip -0.1in
    \centerline{
    \hskip 0.0in
    \includegraphics[width=8.8truecm,angle=0]{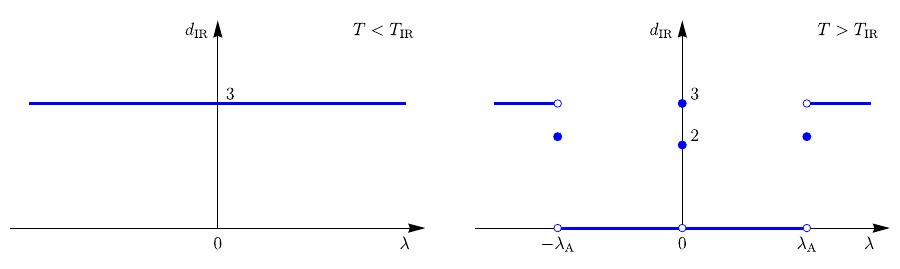}
    }
    \vskip -0.05in
    \caption{Transition to QCD IR phase produces non-analyticity in 
    $d_\fir(\lambda)$, whose generic blueprint is shown on the right.}
    \label{fig:dIR}
    \vskip -0.40in
\end{center}
\end{figure}

We finally wish to convey four important points. 
{\em (i)} Summarizing the accumulated knowledge, we propose that the dimensional 
blueprint $d_\fir(\lambda)$ of QCD in IR phase is shown in Fig.~\ref{fig:dIR} 
(right). At the transition temperature $T_\fir$, this non-analytic behavior 
replaces the constant $d_\fir(\lambda) \!=\! 3$ shown on the left. 
{\em (ii)} Strictly speaking, the extraordinary dimensional transformation at 
$T_\fir$, represented by Fig.~\ref{fig:dIR}, has a robust numerical support in 
pure-glue QCD. However, the results of recent extensive study~\cite{Meng:2023nxf} 
strongly suggest that, at least the structure near $\lambda_\fir \!=\! 0$ is also 
featured in ``real-world" QCD.
{\em (iii)} The previous comment applies to the entire QCD phase diagram in 
Fig.~\ref{fig:Dirac_diagram} (bottom left). While the multidimensional analysis 
has not yet been done in full-QCD case, the fact that the singularity structure at
$\lambda_\fir \!=\! 0$ is preserved~\cite{Meng:2023nxf} strongly suggests that 
the fate of the plateau is similar. Note also that dashing in $T\!=\!T_\fir$ and 
$T\!=\!T_\fuv$ sections of critical lines expresses that these regions were not 
sufficiently studied yet, not even in pure-glue QCD~\cite{Alexandru:2021xoi}.
{\em (iv)} The precise meaning of $d_\fir$ at $\lambda_\fir \!=\! 0$, represented 
in Fig.~\ref{fig:dIR} (right) by two distinct values, is as follows. While 
$d_\fir(0) \!\equiv\! \lim_{L \to \infty} d_\fir(0,L) \!=\!3$ is simply the 
IR dimension of exact zeromodes, 
\begin{equation}
    d_\fir^+(0) \equiv \lim_{\epsilon \to 0} \lim_{L \to \infty} 
    d_\fir(0,\epsilon, L) \approx 2
    \label{eq:325}
\end{equation}
reflects the volume scaling of smallest non-zero modes. In Eq.~\eqref{eq:325} 
we introduced the notation $d_\fir(\lambda_1,\lambda_2,L)$, where 
IR dimension is obtained from average $\efNm$ involving modes from the range 
$\lambda_1 \!<\! \lambda \!<\! \lambda_2$. Surprising results of 
Refs.~\cite{Alexandru:2021pap, Meng:2023nxf}, support the indicated 
$d_\fir^+(0) \!\neq\! d_\fir(0^+)$.

\begin{acknowledgments}
A.A. is supported in part by the U.S. DOE Grant No. DE-FG02-95ER40907.
I.H. acknowledges the discussions with Peter Marko\v{s} and his input
on localization lengths. 
\end{acknowledgments}

%\vfill\eject

%\bibliographystyle{JHEP}
\bibliography{my-references}

\end{document}